\documentclass[useAMS,usenatbib]{mn2e}

\usepackage{times}
\usepackage{amsmath}
\usepackage{graphicx}

\newcommand{\dd}{{\rm d}}
\newcommand{\comovingdist}{D_{\rm C}}

\title[Redshifts and logarithmic wavelength shift]
      {Reinventing the slide rule for redshifts:
         the case for logarithmic wavelength shift}
\author[I.~K.~Baldry]{Ivan K.\ Baldry\\
Astrophysics Research Institute, Liverpool John Moores University, IC2, 
Liverpool Science Park, 146 Brownlow Hill, Liverpool, L3 5RF, UK
}

\begin{document}

\date{2018 December 17th}

\pagerange{\pageref{firstpage}--\pageref{lastpage}} \pubyear{2018}

\maketitle

\label{firstpage}

\begin{abstract}
Redshift is not a shift, it is defined as a fractional change in wavelength. 
Nevertheless, it is a fairly common misconception 
that $\Delta z \, c$ represents a velocity where $\Delta z$ is the redshift
separation between two galaxies. 
When evaluating large changes in a quantity, 
it is often more useful to consider logarithmic differences. 
Defining $\zeta = \ln \lambda_{\rm obs} - \ln \lambda_{\rm em}$
results in a more accurate approximation for line-of-sight velocity and, 
more importantly, 
this means that the cosmological and peculiar velocity terms become additive: 
$\Delta \zeta \, c$ can represent a velocity at any cosmological distance. 
Logarithmic shift $\zeta$, or equivalently $\ln (1+z)$, should arguably 
be used for photometric redshift evaluation. 
For a comparative non-accelerating universe, used in cosmology, 
comoving distance ($\comovingdist$) is proportional to $\zeta$. 
This means that galaxy population distributions in $\zeta$, rather than $z$, 
are close to being evenly distributed in $\comovingdist$, and 
they have a more aesthetic spacing when considering galaxy evolution. 
Some pedagogic notes on these quantities are presented. 
% $z$, $\zeta$, velocity, scalefactor, distance, frames. 
\end{abstract}


\section{Redshift is not a shift}

The definition of redshift is given by 
\begin{equation}
  z \: = \: \frac{\lambda_{\rm obs} - \lambda_{\rm em}}{\lambda_{\rm em}} \mbox{~~~,}
\label{eqn:z-def}
\end{equation}
where $\lambda_{\rm obs}$ is the observed wavelength and $\lambda_{\rm em}$ is
the emitted or rest-frame wavelength 
(e.g.\ eq.~7 of \citealt{HT35}).  For low redshifts, it
is common to quote $z\,c$ for observed galaxies as a recession velocity in
units of $\mathrm{km\,s}^{-1}$.  This is related to the approximation
\begin{equation}
  z_{\rm pec} \: \simeq \: \frac{v}{c}    
\label{eqn:z-pec1}
\end{equation}
where $z_{\rm pec}$ is the redshift (or blueshift) caused by a line-of-sight
peculiar velocity ($v$) component.  This sometimes leads to the {\em
  incorrect} assumption that the `velocity' due to the cosmological expansion
and the peculiar velocity add, or that the redshifts add. 
\citet{DS14} show how, that even at modest redshift, 
the peculiar velocity can be significantly overestimated by 
naively subtracting the cosmological redshift from the observed redshift.

The correct formula for relating redshift terms, 
also incorporating the Sun's peculiar motion, can be given by 
\begin{equation}
  1 + z_{\rm cmb} 
\: = \: (1 + z_{\rm helio}) (1 + z_{\rm pec,\odot}) 
\: = \: (1 + z_{\rm cos}) (1 + z_{\rm pec}) \mbox{~~~,}
\label{eqn:combine-z}
\end{equation}
where $z_{\rm cmb}$ and $z_{\rm helio}$ are the redshifts of an observed
galaxy in the cosmic-microwave-background (CMB) frame 
and heliocentric frame, respectively, 
$z_{\rm pec,\odot}$ is the component caused by the motion of our Sun wrt.\ the
CMB frame toward the observed galaxy, $z_{\rm pec}$ is caused by the peculiar
velocity of the observed galaxy, and $z_{\rm cos}$ is the cosmological
redshift caused by the expansion of the Universe only.  This is evident from
considering the definition of redshift, i.e., {\em `one plus redshift' has a
multiplicative effect on wavelength} \citep{Harrison74}.  
Note there is also a term for gravitational redshift and the heliocentric redshift 
should be determined correctly from the observed redshift.

Taking the difference in redshifts between two galaxies that are at the same distance, 
we obtain 
\begin{equation}
  \begin{split}
  \Delta z \: = \: z_1 - z_2  & = \:  (1 + z_{\rm cos}) \, (z_{\rm 1,pec} - z_{\rm 2,pec}) \\ 
     & \simeq \:  (1 + z_{\rm cos}) \, \frac{v_1 - v_2}{c} 
  \end{split}
  \mbox{~~~,}
 \label{eqn:simplification1}
\end{equation}
using the approximation of Eq.~\ref{eqn:z-pec1}.  So it appears that to
estimate the velocity difference requires knowledge of the cosmological
redshift, though typically one could just set $\Delta v = \Delta z\,c / (1 +
z_1)$, for example, or use one plus the average redshift for the denominator
\citep*{DdZdT80}.  
This is a well known consideration when determining the velocity dispersions
of galaxy clusters. 
A related consequence for counting galaxies in cylinders (e.g.\ \citealt{balogh04})
is that to allow a fixed maximum extent in {\em  velocity} difference 
around a galaxy requires increasing the extent in $\Delta z$ with redshift proportional to $1 + z$. 

Revisiting the approximation, 
the peculiar redshift is accurately given by the Doppler shift formula:
\begin{equation}
  1 + z_{\rm pec} \: = \: \gamma (1+ \beta_{\rm los})
 \label{eqn:doppler}
\end{equation}
where $\gamma = (1-\beta^2)^{-1/2}$ is the Lorentz factor and $\beta_{\rm
  los}$ is the line-of-sight velocity divided by the speed of light.  Using
Taylor series expansion, we can then simplify to:
\begin{equation}
  z_{\rm pec} \: \simeq \: 
  \beta_{\rm los} + \frac{1}{2} \beta^2 + \frac{1}{2} \beta^2 \beta_{\rm los}
  \mbox{~~~.}
\end{equation}
This simplifies further to $z_{\rm pec} \simeq \beta_{\rm los}$ after dropping the
higher order terms.  This is usually sufficiently accurate for use in
astrophysics but it is worth bearing in mind that it is an approximation.

\section{Logarithmic shift zeta}

\begin{figure}
\includegraphics[width=0.48\textwidth]{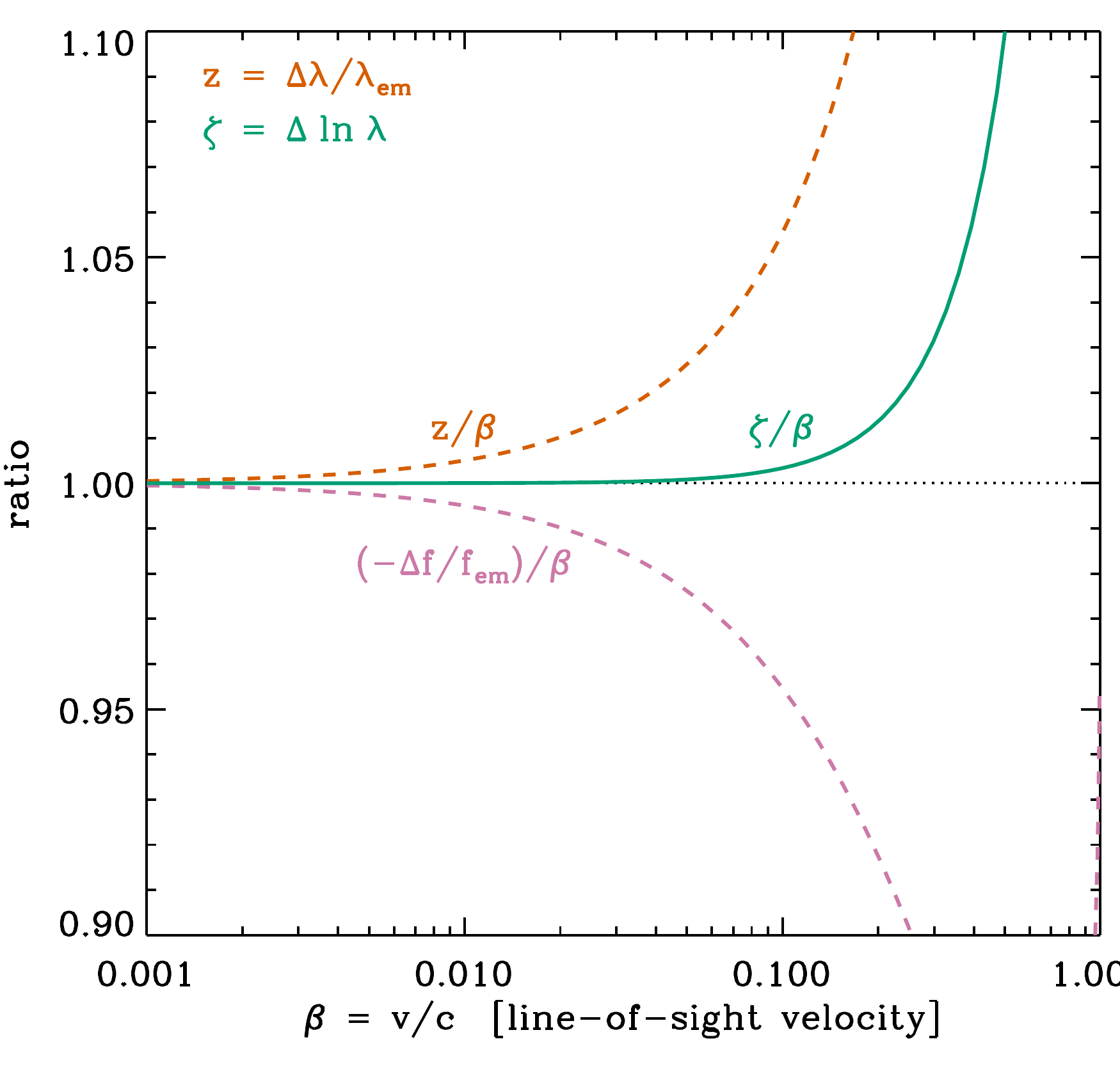}
\caption{Comparison between approximations for recession velocity, i.e.,
assuming pure line-of-sight motion ($\beta_{\rm los} = \beta$). 
The Doppler formula  is used to compute the redshift (Eq.~\ref{eqn:doppler}), 
zeta (Eq.~\ref{eqn:zeta}) 
and the radio definition of velocity as a function of $\beta$.
Notably zeta remains an accurate approximation of recession velocity, 
within a percent, up to $0.1\,c$.}
\label{fig:a}
\end{figure}

Determining redshifts by cross correlation makes it evident 
that a `redshift' or velocity measurement is actually a shift on a logarithmic 
wavelength scale \citep{TD79}. So arguably it is more natural to define a 
quantity (here called zeta) that is a logarithmic shift as 
\begin{equation}
  \zeta \: = \: \ln \lambda_{\rm obs} - \ln \lambda_{\rm em} \: = \: \ln (1+z) \mbox{~~~.}
\label{eqn:zeta}
\end{equation}
First we check its approximation for velocity, using Taylor series, 
\begin{equation}
 \begin{split}
  \zeta_{\rm pec}  & = \: -\frac{1}{2} \ln({1-\beta}^2) + \ln({1+\beta_{\rm los}})  \\
  & \simeq  \: \beta_{\rm los} + \frac{1}{2} (\beta^2 - \beta_{\rm los}^2) +
                                  \frac{1}{3} \beta_{\rm los}^3
 \end{split}
\end{equation}
from the natural logarithm of Eq.~\ref{eqn:doppler}.  Such that $\zeta_{\rm pec}$ 
is always a more accurate approximation for $\beta_{\rm los}$ than $z_{\rm pec}$, 
with the quadratic term vanishing for pure line-of-sight motion.  
Figure~\ref{fig:a} shows a comparison between the redshift, zeta and `radio 
definition' approximations for recession velocity. 

Given the improved accuracy, it is reasonable to use
\begin{equation}
\zeta_{\rm pec} \: \simeq \: \frac{v}{c}
\label{eqn:z-pec2}
\end{equation}
for peculiar velocities. 
This is used implicitly when velocity dispersions of galaxies 
are determined from a logarithmically binned wavelength scale \citep{Simkin74}.

More importantly, the use of zeta means that, the equivalent of Eq.~\ref{eqn:combine-z}
for relating redshift terms becomes 
\begin{equation}
  \zeta_{\rm cmb} 
\: = \: \zeta_{\rm helio} + \zeta_{\rm pec,\odot} 
\: = \: \zeta_{\rm cos}  + \zeta_{\rm pec} \mbox{~~~.}
\label{eqn:additive}
\end{equation}
It is immediately evident that the separation in zeta between two galaxies 
at the same distance is related to velocity directly by 
\begin{equation}
  \Delta \zeta \: \simeq \: \frac{\Delta v}{c}
  \label{eqn:delta-v-zeta}
\end{equation}
with no dependence on the choice of frame or cosmological redshift. 
In addition to being more accurate than Eq.~\ref{eqn:simplification1}, it 
is precisely symmetric when determining the separations in velocity between
two or more galaxies, i.e., there is no need to pick a fiducial redshift. 
A velocity dispersion is given by $\sigma(\zeta)$ regardless of the frame. 

% \subsection{Redshift measurements}

Redshift measurement errors can also be addressed as follows. 
Spectroscopic or photometric redshifts are generally estimated 
by matching a template to a set of observed fluxes at different wavelengths. 
In order to determine the redshift, the template must be shifted in $\ln \lambda$, 
thus we can immediately see that: 
\begin{equation}
  \sigma(\zeta) \: = \: \sigma[\Delta \ln(\lambda)]  \mbox{~~~,}
\end{equation}
which is the uncertainty in the logarithmic shift between the observed and emitted wavelengths. 
Alternatively the redshift uncertainties are often quoted in fractional form: 
\begin{equation}
 \sigma(\zeta) \: \simeq \: \frac{\sigma(z)}{1+z} \mbox{~~~.}
\end{equation}
Either can be related to a velocity uncertainty (Eq.~\ref{eqn:delta-v-zeta}), 
and it is thus reasonable to quote spectroscopic measurements 
using velocity uncertainties \citep{baldry14}. 
The concern is that some papers quote redshift errors in km/s using 
$\sigma(z)\,c$ (e.g.\ \citealt{colless01}),
which does {\em not} represent a physical velocity uncertainty 
even though it has the same units. 

% \subsection{photometric redshifts}

It is appropriate to treat the evaluation of photometric redshift errors in the same way 
and determine the uncertainties in $\zeta$. The typical use of quoting 
${\sigma(z)}/(1+z_{\rm spec})$, where $z_{\rm spec}$ is a spectroscopic redshift, 
for the performance of photometric redshift estimates, approximates this 
(e.g.\ \citealt{brinchmann17}). 
This is somewhat inelegant 
because the uncertainties on photometric redshifts are obtained using
spectroscopic redshifts in the denominator. 
This is no such problem using $\sigma(\zeta)$ and it is more natural since a 
measurement corresponds to a shift in $\ln \lambda$. 
This is just a recognition that fractional differences between two quantities
($1+z$ in this case) depend on a fiducial value 
whereas logarithmic differences are symmetric. 
More importantly, 
this strongly suggests that probability distribution functions, for example, 
should be assessed as a function of zeta 
(binning, outliers, biases, second peak offsets) rather than $z$. 
\citet{Rowan-R03} used $\log_{10}(1+z)$, which equals $\zeta / \ln(10)$,
in his analysis including plots but this is far from standard in the 
literature. 

\section{Cosmological scalefactor}

At a team meeting, I once presented a slide jokingly noting that 
``z is an abomination, it is neither multiplicative, additive or a shift''. 
Of course, redshift's saving grace is that 
a human's computational ability is sufficient to convert 
$z$ to the inverse scalefactor, add unity and you get 
$1 + z_{\rm cos} = a^{-1}$, where $a$ is the cosmological scalefactor
with the common convention that the present-day value $a_0=1$. 

\begin{figure*}
\includegraphics[width=0.9\textwidth]{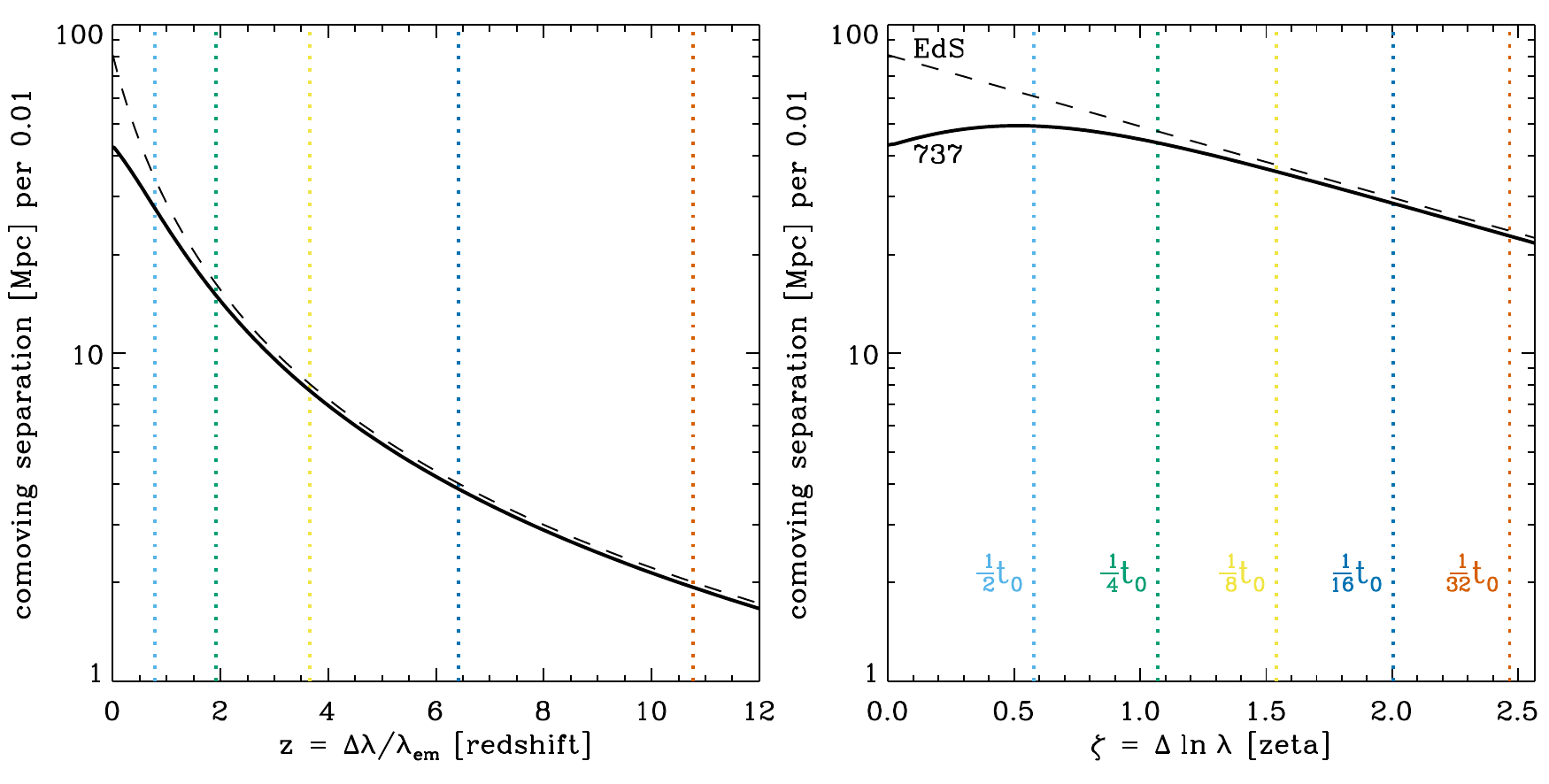}
\caption{Comparison between spacing in redshift and zeta.
The black lines show the comoving separation per 0.01
in $z$ (left) and $\zeta$ (right) (Eq.~\ref{eqn:separations}). 
The solid lines represents the `737 cosmology' 
($h=0.7$, $\Omega_{\rm m}=0.3$, $\Omega_{\Lambda}=0.7$)
while the dashed lines represent an Einstein-de-Sitter cosmology
($h=0.37$ arbitrary, $\Omega_{\rm m}=1$). 
The dotted lines show the points
at which the universe was one half, one quarter, etc., of its present-day 
age for the 737 cosmology.}
\label{fig:b}
\end{figure*}

Using the logarithmic shift $\zeta$, the relationship is
evidently $\zeta_{\rm cos} = \ln a^{-1}$. 
Spacing in logarithm of the scalefactor has desirable properties 
when considering galaxy populations or cosmology (Table~\ref{table}). 
Figure~\ref{fig:b} shows the separation in line-of-sight comoving distance
($\comovingdist$)
versus redshift and zeta for two different cosmologies. 
The black lines show 
\begin{equation}
  S_z \: = \: 0.01 \frac{\dd \comovingdist}{\dd z} \mbox{~~~and~~~}
  S_{\zeta} \: = \: 0.01 \frac{\dd \comovingdist}{\dd \zeta} 
\label{eqn:separations}
\end{equation}
in each plot. 
These are inversely proportional to $\dot{a}/a$ (e.g.\ \citealt{Hogg99}) 
and $\dot{a}$, respectively. Notably $S_{\zeta}$ varies less, particular at $\zeta < 1$. 
This is a desirable property since large-scale structure is evaluated 
using comoving distances. Spacing in $\zeta$ corresponds to
constant velocity and approximately constant comoving distance. 

The turnover in $S_{\zeta}$ demonstrates the onset of dark energy
dominating the dynamics for the `737 cosmology'. 
This is evident even without the comparison to the Einstein-de-Sitter 
(EdS) cosmology because for a non-accelerating universe ($\ddot{a}=0$), 
$S_{\zeta}$ is constant. 
For the EdS model, $S_z \propto a^{3/2}$ and $S_{\zeta} \propto a^{1/2}$
so that 
\begin{equation}
\ln S_{\zeta} = -\frac{1}{2} \zeta + \ln (0.01\,c/H_0) \mbox{~~~,}
\end{equation}
which explains why the dashed line is straight in the right plot of 
Figure~\ref{fig:b}. See, for example, fig.~2 of \citet{aubourg15} for related 
plots [using $\dot{a}$ and $\ln(1+z)$] comparing different 
models of dark energy, and \citet{SR15}
who advocated changing the redshift variable to $\ln(1+z)$
in analysis of luminosity distance residuals. 

Also shown in Figure~\ref{fig:b}, with vertical lines, 
are the points at which the universe halves its age (737 cosmology),
with increasing $z$ and $\zeta$. 
For $z$, the last half of cosmic time covers only a small fraction
of the plot ($z<0.8$), whereas for $\zeta$, the spacing is approximately logarithmic in time. 
For an EdS model, it would be equally spaced in $\ln t$ because $a \propto t^{2/3}$. 
For the 737 cosmology, an increase in $\zeta$ of $\sim 0.5$ corresponds to halving 
the age of the universe across the epochs shown. 
A generic plot related to galaxy evolution
shows the cosmic star-formation rate (SFR) density, logarithmically
scaled, versus $z$ but often scaled linearly in $\ln(1+z)$
\citep{HB06,MD14}. 
This a recognition of the aesthetic of $\ln(a)$ separation.

\begin{table}
\caption{zeta-redshift-scalefactor lookup}
\label{table}
\begin{tabular}{llll} \hline
  $\zeta$ &  $z$ &  $a$  & note\\ \hline
%     0    &  0   &  1    &  \\
     0.1&    0.105&    0.905& $\sim$ present-day galaxy properties \\
     0.5&    0.649&    0.607& $\sim$ transition to cosmic acceleration\\
     1.0&     1.72&    0.368& $\sim$ peak of cosmic SFR density \\
     1.5&     3.48&    0.223&  \\
     2.0&     6.39&    0.135& $\sim$ end of reionization \\
     2.5&     11.2&   0.0821&  \\
     3.0&     19.1&   0.0498& $\sim$ first stars  \\
     7.0&     1096& 0.000912& $\sim$ matter-radiation decoupling\\ \hline
\end{tabular} \\
% Alternative suggestions for zeta symbol: $\hat{z}$, $\mathcal{Z}$ .
\end{table}

\section{Closing remarks and personal comments}

In closing, redshift $z$ started out being considered as a `recession velocity' but 
is now considered as the inverse scalefactor minus unity when assuming $z = z_{cos}$,  
noting also that $z \sim \zeta$ at $z \ll 1$ and $z \sim a^{-1}$ at $z \gg 1$. 
Using the logarithmic shift $\zeta$, 
the cosmological and peculiar velocity terms are additive (Eq.~\ref{eqn:additive}). 
In addition, linear spacing in $\zeta$ corresponds to logarithmic spacing in $a$, 
which is often a practical and aesthetically desirable feature for plots highlighting
cosmological models and galaxy evolution. 
Astronomers regularly use logarithmic differences, magnitude and dex, 
so it would be natural to use logarithmic shift for wavelength. 

Selected points are given below: 
\begin{itemize}
\item Use of $z \, c$ for galaxy recession velocities is poor practice especially 
  beyond a couple of thousand km/s. 
\item Regarding $\zeta \, c$, it is neat that the quadratic term vanishes
  for pure line-of-sight motion. I appreciate this is a special case 
  for peculiar velocities but it is arguably more appropriate for 
  `recession velocity' out to $\zeta \sim 0.1$. 
\item For sources at the same distance, $\Delta z \, c$ is not a velocity,
  $\Delta \zeta \, c$ is a velocity other than for highly relativistic sources. 
\item Use of $\zeta$, or $\ln (1+z)$, is natural for studies that deal 
  with the combination of cosmological and velocity terms. 
\item Photometric redshift analysis should arguably use $\zeta$ as standard
 including presentation and diagnostics. 
 These measurements are effectively analysing shifts in $\ln (\lambda)$. 
\item A plot of $z_{\rm phot}$ versus $z_{\rm spec}$ is inelegant on two counts: 
  it does not relate to the logarithmic shift nature of the measurements, 
  and the spacing is aesthetically poor. 
\item The Hubble-Lema\^{i}tre law $v=H_0 D$ is exact for a 
non-accelerating universe if we use {\em velocity and distance definitions}
$v=\zeta \, c$ and $D = \comovingdist$ (line-of-sight comoving distance).
Thus any deviations from the `law', in this form,
reflect accelerating or decelerating expansion.
\end{itemize}

Comments on the revision history of this paper are given below: 
\begin{itemize}
\item An earlier iteration of this paper was rejected by MNRAS
  (with the title
  ``Shouldn't we be using a shift in logarithmic wavelength as standard?'').
  The anonymous referee noted that it was just an argument for
  ``re-inventing the slide rule'': harsh but fair.
  I have used this quote in the revised title.
\item The same iteration was also rejected as a tutorial by PASP. The
  referee noted ``It isn't exactly a tutorial, ... it is more a plea to
  established astronomers for a revision of notation.
  That notation is so deeply embedded in
  the literature that most working astronomers would not think that
  the small benefits of changing it would be worth the disruption and
  confusion that would result''.  I would argue that confusion, related
  to $z$ and velocity, for example, already exists and will continue;
  I've noticed it many times.
  While the referee's view will be common, 
  I think there are some uses mentioned in this paper where switching to
  $\zeta$, or $\ln(1+z)$ to avoid a new symbol, is more
  readily justified.
\item The tone of the MNRAS submitted version was changed somewhat for arXiv v1,
   along with other minor changes.
\item A reference and note on Hubble-Lema\^{i}tre law were added, following
  comments from W.~Sutherland, for arXiv v2.
\end{itemize}

\bibliographystyle{mn2e-williams}
\bibliography{galaxies,surveys,cosmology}

\label{lastpage}

\end{document}